\documentclass[
reprint,
amsmath,amssymb,
aps,prl
]{revtex4-2}

\usepackage[dvipsnames]{xcolor}

\usepackage{graphicx}
\usepackage{dcolumn}
\usepackage{bm}
\usepackage{color}
\usepackage[unicode]{hyperref}
\hypersetup{
unicode=true,         
plainpages=false,
colorlinks=true,       
linkcolor=blue,          
citecolor=blue,        
urlcolor=blue           
}

\usepackage{ulem}

\begin{document}

\preprint{APS/123-QED}

\title{A Kapitza Trap for Ultracold Atoms}

\author{Jian Jiang, Erik Bernhart, Marvin Röhrle, Jens Benary, Marvin Beck, Christian Baals,}
\author{Herwig Ott}
\email{ott@physik.uni-kl.de}
\affiliation{Department of Physics and Research Center OPTIMAS, Technische 
Universit\"at Kaiserslautern, 67663 Kaiserslautern, Germany}%


\date{\today}

\begin{abstract}
We report on the experimental realization of a Kapitza trap for ultracold atoms. Using time-periodic attractive and repulsive Gaussian potentials, we create an effective trap for ultracold neutral atoms in a regime where the time average of the potential is equal to zero. We analyze the role of experimental imperfections, the stability of the trapped atomic cloud, and the magnitude of the effective potential. We find good agreement with the high-frequency expansion of the underlying system dynamics. Our experimental approach opens up new possibilities to study Floquet systems of neutral atoms. 

\end{abstract}

\maketitle



Manipulating states and dynamics of a system by periodic driving is known as Floquet engineering \cite{Weitenberg2021,Bukov2015}, which is increasingly employed in many areas of physics, including cold atoms \cite{Eckardt2017}, photonics \cite{Ozawa2019}, and solid-state physics \cite{Oka2019}. By designing suitable periodic driving, one can engineer an effective time-independent Hamiltonian with properties that are otherwise not attainable in the corresponding static system. Kapitza's pendulum \cite{Kapitza1951,Landau1976}, an inverted pendulum that is dynamically stabilized by a fast driving of its pivot point, is the most prominent example of such engineering in classical physics. Many applications of Kapitza stabilization in quantum systems have been proposed, including the breaking of translation symmetry \cite{Rajagopal2017}, the periodically driven sine-Gordon model \cite{Citro2015},  the stabilization of bright solitons in a Bose-Einstein condensate (BEC) \cite{Abdullaev2003}, cold atoms with oscillating interactions \cite{Abdullaev2003pra}, optical molasses \cite{Bagnato94}, preparation of molecular ions \cite{Smirnova2003}, the stability of optical resonators \cite{Torosov2013}, polariton Rabi oscillation \cite{Voronova2016}, and unconventional dynamical phases \cite{Lerose2019}. In particular, Kapitza stabilization can be employed to trap particles. The most notable example of such an application is the Paul trap \cite{Paul1990,Goldman2014}, where a saddle point potential is modulated periodically to create a confining harmonic potential. Light confinement in dielectric structures with a transverse refractive index distribution periodically modulated in the longitudinal coordinate has been proposed \cite{Alberucci2013} and experimentally demonstrated\cite{Muniz19}. Regarding trapping atoms by laser fields, Ref. \cite{Martin2018} has proposed and analyzed how a spatially oscillating red-detuned optical lattice could localize a repulsively interacting BEC, and Ref. \cite{Ido2003} has investigated the case where a time-periodic localized attractive and repulsive Gaussian potential with a vanishing time average creates a conservative trapping potential. 

In this letter, inspired by the theoretical work in Ref. \cite{Ido2003}, we demonstrate a "Kapitza trap" for ultracold atoms and thereby extend the field of Kapitza pendulum physics towards the regime of ultracold quantum gases. The Kapitza trap is created by superimposing time-modulated focused laser beams to induce a time-varying dipole trapping potential for neutral atoms. 
\begin{figure}[htbp]
\centering
\includegraphics[width=8.6cm]{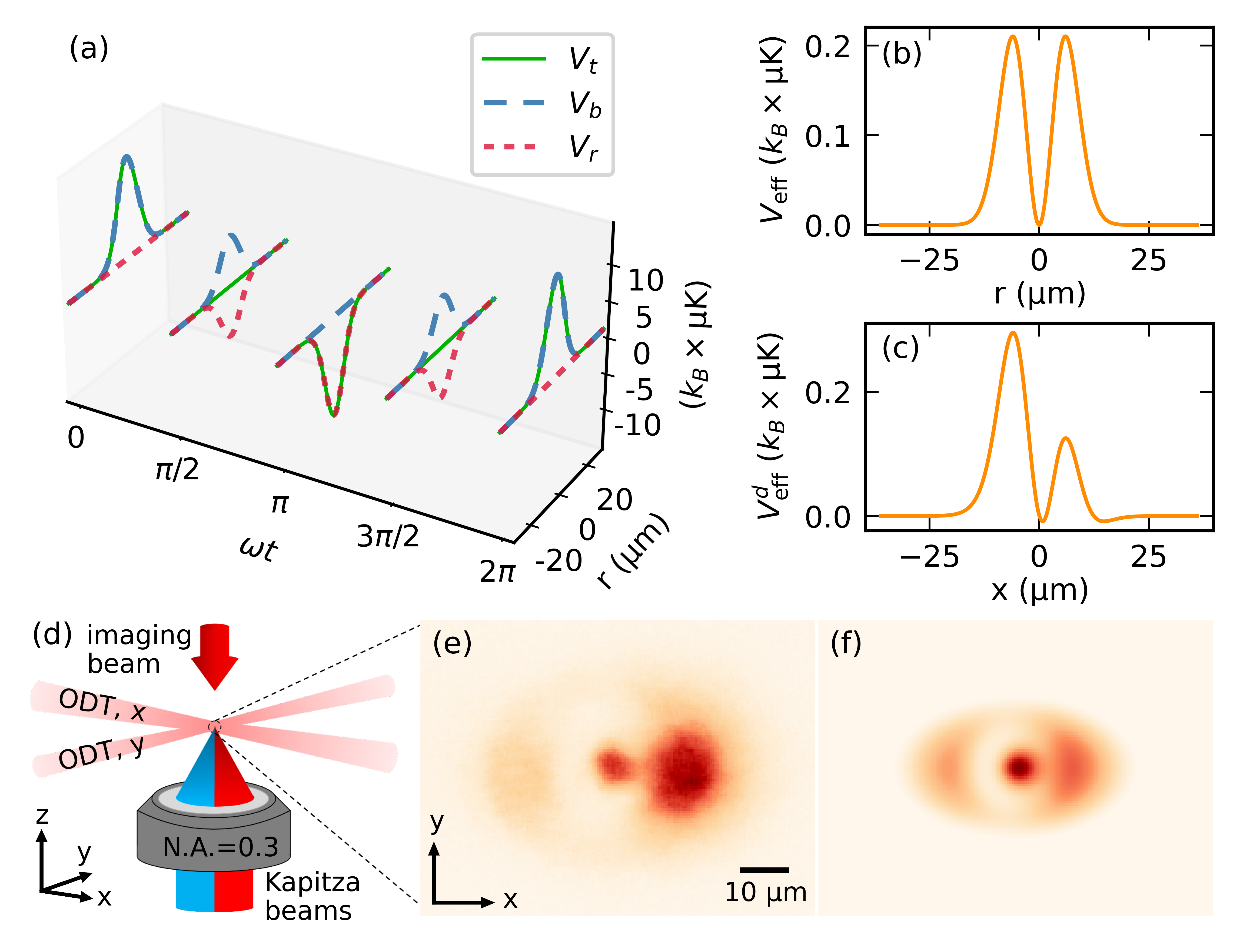}
\caption{Working principle of the Kapitza trap. Kapitza laser: waist $\sigma=12\,\rm{\mu m}$, amplitude $V_{0}=k_B\times 14\,\rm{\mu K}$, modulation frequency $\omega/2\pi=2.4\,\rm{kHz}$. (a) Time-modulated Gaussian potential $V_{t}$ for one oscillation period. $V_{b}$ and $V_{r}$ are the individual potentials of the two Kapitza lasers. (b) Effective potential $V_{\rm{eff}}$ in the high frequency limit [Eq.\,(\ref{eq:VeffG})]. The distance between the two peaks is $\sigma$. (c) Effective potential $V_{\rm{eff}}^d$ at $y=0$ for a small displacement of $d=120\,$nm between the two laser beams [Eq.\,(\ref{eq:Veffk})]. (d) Experimental setup. (e) Absorption image of the atomic cloud in the Kapitza trap. (f) Calculated atomic density plotted with the same coordinate and length scale of (e). Densities in both cases are rescaled from 0 to 1, where 1 is the maximum density in each individual case.}
\label{fig:1}
\end{figure}

Specifically, we confine ultracold $^{87}\rm{Rb}$ atoms using a time-periodic Gaussian potential of the form [Fig. \ref{fig:1}(a)],
\begin{equation}
V_{t}\left(r,t\right)
=V_{0}\,\mathrm{exp}\left(-\frac{2r^2}{\sigma^2}\right)\mathrm{cos}\left(\omega t\right)
\label{eq:V1G}
\end{equation}
where $\sigma$ and $V_{0}$ are the waist and amplitude of the Gaussian profile respectively, and $r=(x^2+y^2)^{1/2}$ is the radial coordinate. In practice, $V_{t}$ is the result of the superposition of a repulsive potential
\begin{align}
V_{b}\left(r,t\right) &=
V_{b0}\,\mathrm{exp}\left(-\frac{2r^2}{\sigma_b^2}\right)\frac{1+\mathrm{cos}\left(\omega t\right)}{2} \\
&= G_b(r)\left[1+\mathrm{cos}\left(\omega t\right)\right]/2
\label{eq:Vb}
\end{align}
and an attractive potential
\begin{align}
V_{r}\left(r,t\right) &=
V_{r0}\,\mathrm{exp}\left(-\frac{2r^2}{\sigma_r^2}\right)\frac{1-\mathrm{cos}\left(\omega t\right)}{2} \\
&= G_r(r)\left[1-\mathrm{cos}\left(\omega t\right)\right]/2
\label{eq:Vr}
\end{align}
created by a blue-detuned laser (777.827 nm) and a red-detuned laser (782.325 nm) \cite{SM1}. These two lasers are referred to as Kapitza lasers. In Eq.\,(\ref{eq:Vb}) and Eq.\,(\ref{eq:Vr}), we have introduced $G_b$ and $G_r$ to represent their unmodulated Gaussian profiles. For equal power, the two laser beams induce dipole potentials with the same magnitude but opposite sign for $^{87}\rm{Rb}$, i.e., $V_{b0}=-V_{r0}=V_{0}$. Both lasers have the same linear polarization, but their power modulation is out of phase by 180$^\circ$. In order to ensure maximum spatial overlap and equal waists ($\sigma_b=\sigma_r=\sigma$) of the two beams at the position of the atoms, we couple both lasers into the same polarization-maintaining single-mode fiber. The resulting effective potential $V_{\rm{eff}}$ in the high-frequency limit then reads to first order \cite{Ido2003,Rahav2003} [Fig. \ref{fig:1}(b)] 
\begin{equation}
V_{\rm{eff}}\left(r\right)
=\frac{4V_0^2}{m\sigma^4\omega^2}\,r^2\,\mathrm{exp}\left(-\frac{4r^2}{\sigma^2}\right)\,,
\label{eq:VeffG}
\end{equation}
where $m$ is the mass of a trapped atom. Applying the harmonic approximation to $V_{\rm{eff}}$ at $r=0$, we derive the oscillation frequency $\Omega=2\sqrt{2}V_0/m \sigma^2\omega$ of an atom in the Kapitza trap. Eq.\,(\ref{eq:VeffG}) is a good approximation for $\omega\gg\Omega$ \cite{Ido2003}. Because $V_{\rm{eff}}$ stems from the mean kinetic energy stored in the micromotion of the particles, it is always positive and peaks at maxima of the kinetic energy of the particle's micromotion. The same microscopic dynamics is also responsible for the effective potential of a Paul trap. 

The two maxima of $V_{\rm{eff}}$, $V_{\rm{eff}}^{\max}=V_{\rm{eff}}(r=\pm \sigma/2)$, are separated by the waist of the Kapitza beam, $\sigma$. Note that in order to create the same trapping potential exclusively using lasers without time modulation, one requires a waist that is smaller by a factor of $\sqrt{2}$. While more subtle methods \cite{Ge2020,Wang2018} are available for creating optical potentials with even better resolutions, our approach is, at least in principle, a straightforward way to surpass the diffraction limit modestly. 

The experimental setup is sketched in Fig.\,\ref{fig:1}(d). We start by producing a BEC of $^{87}\rm{Rb}$ atoms confined in a crossed optical dipole trap (ODT, $\lambda=1064$\,nm). The trap frequencies are $\omega_x/2\pi=55\,\mathrm{Hz}$, $\omega_y/2\pi=93\,\rm{Hz}$, and $\omega_z/2\pi=108\,\rm{Hz}$, where the $z$-direction is anti-parallel to the direction of gravity. The central chemical potential of the condensate is $\mu\approx k_B \times 300\,\rm{nK}$. The Kapitza lasers, which are focused on the condensate along the $z$-direction by an objective with a numerical aperture (N.A.) of 0.3, are then linearly ramped up \cite{ramp} and held until the system is probed by in situ absorption imaging using the same objective. Power modulation of the Kapitza lasers is always on during the ramp and the hold time. Since the dominant confinement along the $z$-direction is from the ODT, the total trapping potential is well approximated by the two-dimensional Kapitza trap in the radial direction [Eq. (\ref{eq:VeffG})] and the superimposed ODT.

In Fig.\,\ref{fig:1}(e), we show a typical absorption image of the trapped atomic cloud with Kapitza beams whose waists are 12 $\rm{\mu m}$. The Kapitza trap in the center is clearly visible and demonstrates the basic working principle. The annular depletion zone between the Kapitza trap and the outer ring of atoms is a direct consequence of the rotationally symmetric effective potential, which creates a ring barrier. 

Although both Kapitza beams share the same spatial mode, the different wavelength is likely to cause a slight displacement between their beam profiles via dispersive optical elements. Due to the displacement, the Kapitza trap is always closer to the right side of the outer ring, where the atomic density is higher than the rest of the ring. To quantitatively analyze this phenomenon, we introduce a displacement $d$ along the $x$-axis of the red-detuned beam:
\begin{equation}
G_{r}\left(x,y,d\right)
=V_{r0}\,\mathrm{exp}\left(-\frac{\left(x-d\right)^2+y^2}{\sigma_r^2/2}\right).
\label{eq:Gr}
\end{equation}
The effective potential can then be written as \cite{Rahav2003}
\begin{align}
&V_{\rm{eff}}^{d}\left(x,y,d\right)=\frac{G_{b}+G_{r}}{2}+\frac{1}{m\sigma^4\omega^2}\left\{(x^2+y^2)G_{b}^2+
\notag\right.\\
&\phantom{=\;\;}
\left.2\left[x(x-d)+y^2\right]G_{b} G_{r}+\left[(x-d)^2+y^2\right]G_{r}^2\right\} \notag\\
&\hspace{5em} \equiv V_{\rm res}+V_{\rm K}\,,
\label{eq:Veffk}
\end{align}
where $V_{\rm res}$ is the residual potential induced by the displacement, and $V_{\rm K}$ is the potential of the Kapitza trap, a ring barrier potential similar to $V_{\rm eff}$ but not centered at the coordinate origin when $d\neq 0$. According to Eq. (\ref{eq:Veffk}), a small displacement of 120\,nm is already sufficient to imbalance the Kapitza trap [Fig.\,\ref{fig:1}(c)]. For $d>240\,$nm, which amounts to only 2\,\% of the beam waist, the trap disappears as $V_{\rm res}$ dominates. We determine the displacement as 120 nm (corresponding to $0.01\times\sigma$ by comparing the experimental result and the calculated atomic density \cite{SM2}, which is given by the time-independent Gross-Pitaevskii equation in Thomas-Fermi approximation). Fig.\,\ref{fig:1}(f) shows the reconstructed density for $V_{\rm{eff}}^d (x,y,d=120\,\rm{nm})$ and a chemical potential of $\mu_d=k_B\times 400\,\rm{nK}$. This model captures most of the features observed in the experiment, including the asymmetry of the cloud. 

Similar to the spatial displacement, any other imperfect alignment (e.g., power imbalance, beam waist mismatch) of the two Kapitza beams will result in a finite $V_{\rm res}$. The maximum of $V_{\rm K}$ has the same order of magnitude as $V_{\rm{eff}}^{\max}$. From Eq.\,(\ref{eq:VeffG}), one can deduce that
$V_0/V_{\rm{eff}}^{\max}
=(2\sqrt{2}\,e)(\omega/\Omega) \simeq 8\,\omega/\Omega$. In the experiment shown in Fig.\,\ref{fig:1}(e),  we have $\omega/\Omega=9$, which corresponds to $V_0/V_{\rm{eff}}^{\max} \simeq 72$. Such conditions require that $V_{\rm res}$ is about two orders of magnitude smaller than $V_0$ in order to observe the Kapitza trap. This can only be achieved by a careful alignment of the two laser beams. In the experiment, we change the relative power of the two beams in sub-percent steps until the Kapitza trap appears together with the outer ring. We refer to this as the balanced situation, going away from which the feature disappears already for a few percent beam imbalance.

It is instructive to compare the Kapitza trap for neutral atoms with the Paul trap for charged particles. While the electric field couples directly to the charge of the particles, neutral atoms couple via their polarizability to the light field. Accordingly, the later coupling is orders of magnitude weaker, so realizing a Kapitza trap for neutral atoms is much more challenging. On the other hand, a maximum of the light field can easily be created in free space. Thus, in principle, the Kapitza trap provides a way to confine neutral atoms in all three spatial dimensions by oscillating optical potentials alone without requiring close-by electrodes as in a Paul trap.

We now turn to the stability analysis and characteristics of the Kapitza trap. In the high-frequency limit, $V_{\rm eff}$ remains constant when proportionally increasing $V_0$ and $\omega$ (Eq.\,(\ref{eq:VeffG})). Hence, the trap frequency $\Omega$ also remains unaffected. The ratio $\omega/\Omega$, however, increases. This way, we can test the system behavior in the high- and low-frequency limits by measuring the lifetime of the trapped atoms. This works, however, only in a limited parameter range for a given displacement. While $V_{\rm K}$ will remain unchanged, $V_{\rm res}$ will increase monotonously and therefore become dominant. In other words, a given displacement sets an upper limit for $\omega/\Omega$ in such a test. The highest $\omega/\Omega$ we achieve for a beam waist of 12\,$\rm{\mu m}$ is about 9 (Fig.\,\ref{fig:1}(e)).

\begin{figure}[htbp]
\includegraphics[width=7.5cm]{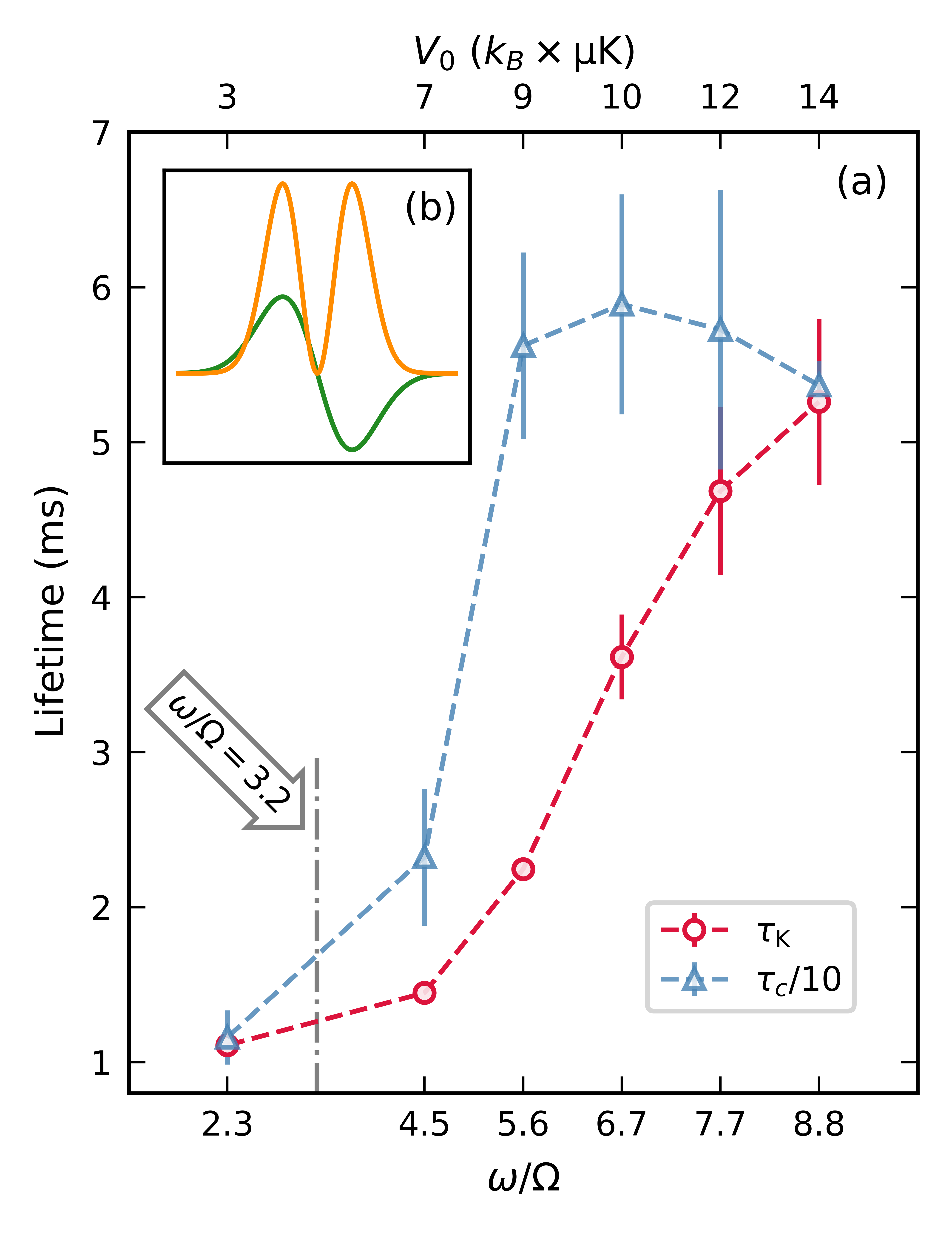}
\caption{(a) Lifetimes of atoms inside the Kapitza trap ($\tau_{\rm{K}}$) and the outer ring ($\tau_c$, scaled down by a factor of ten) in dependence of the modulation frequency $\omega$ for approximately constant $\Omega$, plotted against $\omega/\Omega$. To determine the number of atoms in the Kapitza trap, we integrate the atomic density over an circle with a diameter of $2\sigma/3$ ($8\,\mu m$, see also Fig.\,\ref{fig:1}b). The upper axis shows the potential $V_0$, which is increased proportional to $\omega$. Data points are connected by dashed lines for better visualization. (b) $V_{\rm res}$ (green) and $V_{\rm K}$ (orange) at $\omega/\Omega=8.8$, plotted with the same scale.}
\label{fig:2}
\end{figure}

Fig.\,\ref{fig:2}(a) shows the measured lifetimes of atoms inside the Kapitza trap ($\tau_\text{K}$) and the outer ring ($\tau_c$) as $V_0$ and $\omega$ increase proportionally, keeping $\Omega$ approximately constant \cite{Omega}. Eq.\,(\ref{eq:VeffG}) becomes a more accurate approximation for increasing modulation frequency $\omega$. At the same time, the excitation of the atomic motion is expected to become less pronounced as the time-varying potential couples less efficiently to the atoms at higher frequencies. The above trends are clearly observed in our experiment as measured lifetimes, $\tau_{\rm{K}}$ and $\tau_c$, increase with the modulation frequencies. On the other hand, $\tau_{\rm{K}}$ is still strongly affected by the time-periodic potential even at high frequencies, and it is at least one order of magnitude smaller than $\tau_{\rm c}$ for the same $\omega/\Omega$. This can be attributed to several effects. As shown in Fig.\,\ref{fig:2}(b), $V_{\rm res}$ resulting from the beam displacement pushes atoms away from the trap center. These atoms move in an area with larger oscillating potential, which leads to additional heating. This phenomenon can be analogized to the well-known excess micromotion of charged particles in a Paul trap without proper DC compensation voltages. These atoms move in a region with much larger oscillating potential, which leads to additional heating because the energy of the micromotion can be converted into the energy of the macromotion \cite{Berkeland1998}. Collisions between atoms intensify such a heating process \cite{Bluemel1995}. Light scattering from the Kapitza beams, unstable orbits due to the non-parabolic double well structure of the effective potential and three-body losses constitute additional heating and loss processes, which further reduce the lifetime.

The stability of the Kapitza trap can also be analyzed in terms of the classical equation of motion for the atoms in the trap. Approximating the center of the Kapitza trap with a harmonic trap, the equation of motion can be written as
\begin{equation}
\frac{d^2x}{d\tau^2} - 2q\,\text{cos}(2\tau) x=0
\label{eq:mathieu}
\end{equation}
where $\tau=\omega t/2$ and
\begin{equation}
q=\frac{8V_0}{m\sigma^2\omega^2}=\frac{2\sqrt{2}}{\omega/\Omega}\,.
\label{eq:q}
\end{equation}
Eq. (\ref{eq:mathieu}) is a standard Mathieu differential equation. For $q\le 0.9$, i.e., $\omega/\Omega\ge 3.2$, this equation has a periodic solution, which implies that the atoms undergo closed orbits and are therefore trapped \cite{Foot2005}. In other words, when considering the single-particle picture, $V_t$ can be replaced by $V_{\rm{eff}}$ for $\omega/\Omega\ge 3.2$.

For values of $\omega/\Omega$ larger than 5, $\tau_c$ is roughly on a plateau. Without the Kapitza beams, the atomic density amounts to $3\times 10^14 \rm cm^{-3}$ in the trap center, corresponding to a lifetime of 2\,s. When switched on, the Kapitza trap compresses the outer ring of atoms due an to the effective repulsion and leads to enhanced three-body losses. As they depend on $V_0$ (see Fig. 4), which changes slightly for $\omega/\Omega>5$, one would expect $\tau_c$ are roughly on a plateau. Furthermore, the measured lifetime of several tens of milliseconds could be realistic, if one takes into account that non-condensed atoms have a 6-fold higher three-body loss rate compared to Bose-condensed atoms \cite{Burt1997}.

\begin{figure}[htbp]
\includegraphics[width=8.6cm]{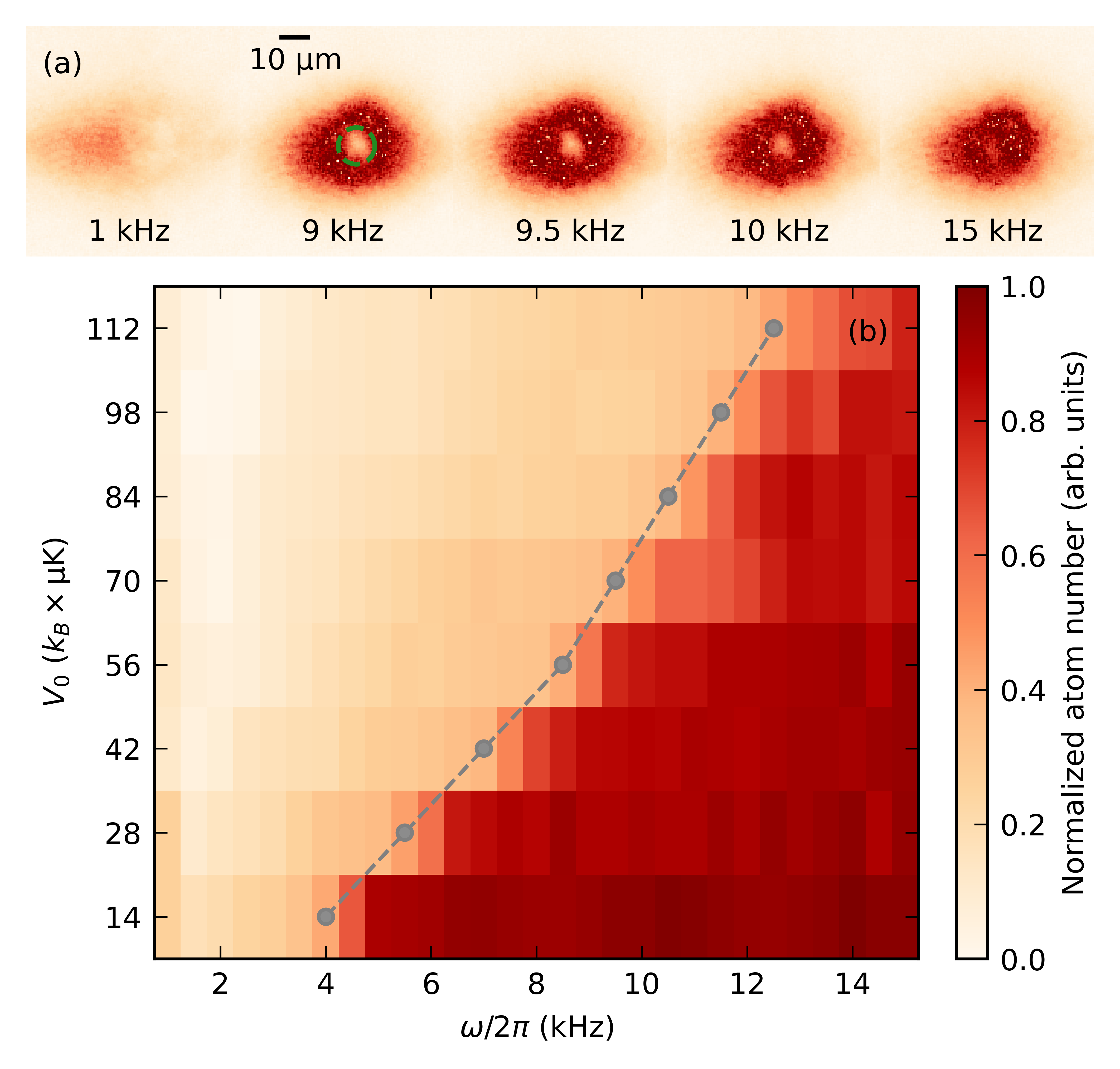}
\caption{(a) Appearance of a hole in the atomic cloud upon applying the Kapitza lasers ($\sigma=6\,\rm{\mu m}$, $V_0=70\,\rm{\mu K}$ ). The dashed green circle defines the hole area, in which we determine the number of expelled atoms. (b) Atom number inside the hole area. Measured atom numbers are normalized to the maximum number of atoms measured inside the green circle. Grey points (with dashed line as a guide to the eye) represent $(-V_{r0},\,\omega/2\pi)$ for forming evident holes (calculated with $V_{b0}=-96\%\,$ and $V_{r0}=96\%\,V_0$).}
\label{fig:3}
\end{figure}

To understand the effect of the beam size on the Kapitza trap, we have also performed experiments with smaller Kapitza beams ($\sigma=6\,\rm{\mu}$m). We first note that the Kapitza trap is no longer visible in the absorption images, which is most likely due to a more significant relative displacement of both beams. However, we observe the appearance of a hole in the density distribution as a signature of the time-modulated potential. The effective potential can be estimated by analyzing the characteristics of this hole. Fig. \ref{fig:3}(a) shows the appearance of the hole when the modulation frequency decreases. At very high modulation frequencies, the effective potential is negligible compared to the chemical potential of the condensate, and only a minor effect is visible. As the modulation frequency decreases, a hole starts to show up at the center of the cloud when the effective potential becomes comparable with the chemical potential. At small modulation frequencies, the effective potential is no longer a good approximation, and the small frequency perturbations strongly affect the cloud. 

For a quantitative evaluation, we take images of the cloud for different combinations of the Kapitza lasers' powers and modulation frequencies, and determine the number of atoms inside the hole area (green dashed circle in Fig. \ref{fig:3}(a)). The result is shown in Fig. \ref{fig:3}(b). We observe that there is a frequency above which the hole disappears for each power level. The transition is rather steep. In order to provide a theoretical estimate for the creation of the hole, we proceed as follows. The effective potential leads to a spatially varying reduction of the density. Averaging the effective potential in the transverse direction over an area A gives
$V_{\mathrm{av}}=A^{-1}\iint dx dy V_{\rm{eff}}^d(x,y,d)$. We chose $\sigma/ \sqrt{2}$ as the radius for the area. An evident hole is then formed, when this averaged potential is equal to the chemical potential of the atoms. These points are shown as grey dots in Fig. \ref{fig:3}(b). The quantitative agreement confirms that the effective potential is a good description for this parameter range. 

\begin{figure}[htbp]
\includegraphics[width=7.5cm]{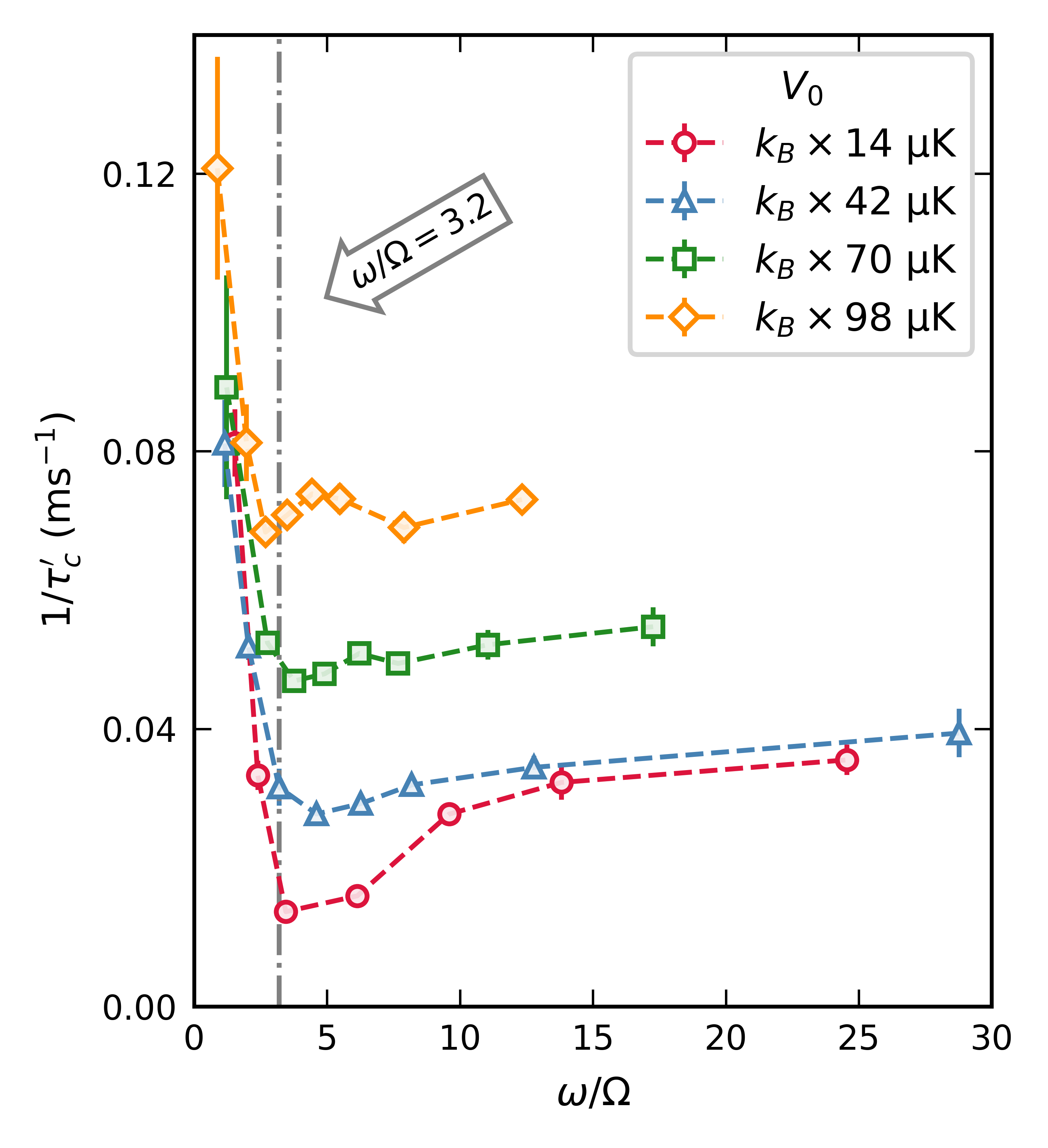}
\caption{Lifetime ($\tau_c^{\prime}$) of the whole atomic cloud in the presence of the Kapitza beams for different strengths. For $\omega/\Omega>3.2$, $\tau_c^{\prime}$ is roughly a constant. Dashed lines are guides to the eye.}
\label{fig:4}
\end{figure}

We measure the lifetime $\tau_c^{\prime}$ of the whole atomic cloud to investigate further the validity of the effective potential approximation for the smaller Kapitza beams. The result is plotted as $1/\tau_c^{\prime}$ against $\omega/\Omega$ [Fig.\,\ref{fig:4}] for better visualization. As one would expect, $\tau_c^{\prime}$ has a similar behavior as $\tau_c$. It first increases and then reaches a plateau beyond a specific $\omega/\Omega$. For larger $V_0$, the compression of the atomic cloud caused by
the repulsion of the effective potential is stronger. Therefore, the corresponding $\tau_c^{\prime}$ on the plateau is smaller.

In summary, we have experimentally demonstrated a Kapitza trap for ultracold atoms. Employing a rapid time-periodic Gaussian potential, realized with red- and blue-detuned focused laser beams, we confine ultracold $^{87}\rm{Rb}$ atoms with an effective ring barrier potential. We validate the approximation of the effective potential in the high-frequency limit. At present, our trapped samples suffer from limited lifetimes, limited achievable modulation frequency, and residual static potentials due to relative beam displacement. Further eliminating this 1$\%$ beam displacement is the key to verifying and sorting out these issues. We suspect the in-vacuum objective used in this experiment induces the displacement since it is designed for monochromatic light. In principle, exclusively using reflective optics after the fiber employed for overlapping the Kapitza beams can avoid chromatic aberration so that the two beams fully overlap and have the same size at the position of the atoms.

Once the lifetime of the atoms in the Kapitza trap is increased, our approach allows for various new research directions. First, we recall that the effective potential exhibits a tunneling barrier. This feature allows for studying quantum tunneling in time-periodic potentials, where the tunneling rate can be reduced or even completely suppressed \cite{Grossmann1991,Eckardt2005,Lignier2007}. A further extension of our work is the detailed study of the low-frequency limit. This is the realm of resonant Floquet systems, where the modulation frequency matches an intrinsic frequency of the system. Such driving has been proposed, for instance, to implement energy and spin filters \cite{Thuberg2017,Reyes2017}  and has recently been extended to interacting systems \cite{Fazzini2021}. Due to the Gaussian envelope of the time-periodic potential $V_t$ [Eq. (\ref{eq:V1G})], our system is intrinsically nonlinear. This makes it also a potential tool for investigating chaos in the classical and quantum regime \cite{DAlessio2016}. 

We gratefully acknowledge discussions with Ido Gilary, Serena Fazzini, and Junhui Zheng. We acknowledge financial support by the Deutsche Forschungsgemeinschaft within the Forschungsgroßgeräteprogramm INST 248/250-1 and the collaborative research center TRR 185, project B3 (number 277625399).

\nocite{*}

\bibliography{Kapitza}

\begin{thebibliography}{43}
\bibitem [{ram()}]{ramp}%
  \BibitemOpen
  \href@noop {} {\bibinfo  {journal} {The ramp durations are 3\,ms and 5\,ms
  for experiments using the bigger ($\sigma=12\,\mu $m) and the smaller
  ($\sigma=6\,\mu$m) Kapitza beams, respectively. The hold time for the results
  shown in Fig. \ref{fig:3}(b) is 5\,ms}\ }\BibitemShut {NoStop}%
\bibitem [{Ome()}]{Omega}%
  \BibitemOpen
\bibfield  {journal} {  }\href@noop {} {\bibinfo  {journal} {Because of the
  displacement between the two Kapitza beams, $\Omega$ cannot be a constant.
  However, $\Omega$ is treated as a constant since it changes less than 2$\%$
  during the measurement.}\ }\BibitemShut {NoStop}%
\bibitem [{blu()}]{blue_residual}%
  \BibitemOpen
  \href@noop {} {\bibinfo  {journal} {The residual potential induced by the
  beam displacement also includes a repulsive part, which can expel atoms from
  the trap. However, the repulsive part only causes a minor effect compared to
  the attractive part. In our system, a 10 $\mu$W 777.827\,nm laser beam with a
  waist of 6\,$\mu$m (i.e., $V_{b0}=k_B\times\,1.4 \rm{\mu K}$) alone shortens
  the lifetime of the BEC from 5\,s to 1\,s}\ }\BibitemShut {NoStop}%
\end{thebibliography}

\end{document}